\documentclass[10pt]{article}
\usepackage{color,soul}

\usepackage{amsmath}
\usepackage{amssymb}

\usepackage{graphicx}

\usepackage{cite}

\usepackage{color} 

\usepackage{setspace} 

\usepackage[labelfont=bf,labelsep=period,justification=raggedright]{caption}

\makeatletter
\renewcommand{\@biblabel}[1]{\quad#1.}
\makeatother

\date{}

\newcommand{\uli}[1]{\smash{\underline{#1}}}
\newcommand{\dund}[1]{\smash{\underline{\underline{#1}}}}

\begin{document}

\begin{flushleft}
{\Large
\textbf{Mechanical cell-matrix feedback explains pairwise and collective endothelial cell behavior in vitro}
}
\\
Ren{\'e} F. M. van Oers$^{1,2,\ast,\ast\ast}$, 
Elisabeth G. Rens$^{1,2,\ast}$,
Danielle J. LaValley$^3$,
Cynthia A. Reinhart-King$^3$,
Roeland M. H. Merks$^{1,2,4,\ast\ast\ast}$
\\
{\bf 1} Life Sciences group, Centrum Wiskunde \& Informatica, Amsterdam, The Netherlands
\\
{\bf 2} Netherlands Consortium for System Biology - Netherlands Institute for Systems Biology, Amsterdam, The Netherlands
\\
{\bf 3} Department of Biomedical Engineering, Cornell University, Ithaca, NY, USA
\\
{\bf 4} Mathematical Institute, Leiden University, Leiden, The Netherlands
\\
$\ast$ Authors contributed equally
\\
$\ast\ast$ Present address: Oral Cell Biology, Academic Centre for Dentistry Amsterdam (ACTA), The Netherlands
\\
$\ast\ast\ast$ E-mail: roeland.merks@cwi.nl
\end{flushleft}

\section*{Abstract}
In vitro cultures of endothelial cells are a widely used model system of the collective behavior of endothelial cells during vasculogenesis and angiogenesis. When seeded in an extracellular matrix, endothelial cells can form blood vessel-like structures, including vascular networks and sprouts. Endothelial morphogenesis depends on a large number of chemical and mechanical factors, including the compliancy of the extracellular matrix, the available growth factors, the adhesion of cells to the extracellular matrix, cell-cell signaling, etc. Although various computational models have been proposed to explain the role of each of these biochemical and biomechanical effects, the understanding of the mechanisms underlying in vitro angiogenesis is still incomplete. Most explanations focus on predicting the whole vascular network or
sprout from the underlying cell behavior, and do not check if the same model also correctly captures the intermediate scale: the pairwise cell-cell interactions or single cell responses to ECM mechanics. Here we show, using a hybrid cellular Potts and finite element computational model,  that a single set of biologically plausible rules describing (a) the contractile forces that endothelial cells exert on the ECM, (b) the resulting strains in the extracellular matrix, and (c) the cellular response to the strains, suffices for reproducing the behavior of individual endothelial cells and the interactions of endothelial cell pairs in compliant matrices. With the same set of rules, the model also reproduces network formation from scattered cells, and sprouting from endothelial spheroids. Combining the present mechanical model with aspects of previously proposed mechanical and chemical models may lead to a more complete understanding of in vitro angiogenesis. 

\section*{Author Summary}
During the embryonic development of multicellular organisms, millions of cells cooperatively build structured tissues, organs and whole organisms, a process called morphogenesis. How the behavior of so many cells is coordinated to produce complex structures is still incompletely understood. Most biomedical research focuses on the molecular signals that cells exchange with one another. It has now become clear that cells also communicate biomechanically during morphogenesis. In cell cultures, endothelial cells---the Òbuilding blocksÓ of blood vessels---can organize into structures resembling networks of capillaries. Experimental work has shown that the endothelial cells pull onto the protein gel that they live in, called the extracellular matrix. On sufficiently compliant matrices, the strains resulting from these cellular pulling forces slow down and reorient adjacent cells. Here we propose a new computational model to show that this simple form of mechanical cell-cell communication suffices for reproducing the formation of blood vessel-like structures in cell cultures. These findings advance our understanding of biomechanical signaling during morphogenesis, and introduce a new set of computational tools for modeling mechanical interactions between cells and the extracellular matrix.

\section*{Introduction}

How the behavior of cells in a multicellular organism is coordinated to form structured tissues, organs and whole organisms, is a central question in developmental biology. Keys to answering this question are chemical and mechanical cell-cell communication and the biophysics of self-organization. Cells exchange information by means of diffusing molecular signals, and by membrane-bound molecular signals for which direct cell-cell contact is required. In general, these developmental signals are short-lived and move over short distances. The extracellular matrix (ECM), the jelly or hard materials that cells secrete, provides the micro-environment the cells live in. Apart from its supportive function, the ECM mediates molecular \cite{Hynes:2009jo} and biomechanical \cite{ReinhartKing:2008cv} signals between cells. Mechanical signals, in the form of tissue strains and stresses to which cells respond \cite{Mammoto:2009bz}, can act over long distances and integrate mechanical information over the whole tissue \cite{Nelson:2005dd}, and also mediate short-range, mechanical cell-cell communication \cite{ReinhartKing:2008cv}. How such mechanical cell-cell communication via the ECM can coordinate the self-organization of cells into tissues is still poorly understood. Here we propose a cell-based model of endothelial cell motility on compliant matrices to address this problem.

A widely used approach to study the role of cell-ECM interactions in coordinating collective cell behavior is to isolate cells (e.g., endothelial cells isolate from bovine aortae or from human umbilical cords or foreskins) and culture them on top of or inside an artificial or natural ECM (e.g., Matrigel). This makes it possible to study the intrinsic ability of cells to form tissues in absence of potential organizing signals or pre-patterns from adjacent tissues. A problem particularly well-studied in cell cultures is the ability of endothelial cells to form blood vessel-like structures, including the formation of vascular-like networks from dispersed cells and the sprouting of spheroids. To this end, cell cultures can be initialized with a dispersion of endothelial cells on top of an ECM material (e.g., Matrigel, collagen, or fibrin) \cite{Folkman:1980vf,Califano:2008ct},  with endothelial spheroids embedded within the ECM \cite{Korff:1999ty,Kniazeva:2009jm}, or with confluent endothelial monolayers  \cite{Vernon.ajp95,Vernon:1995bj,Koolwijk:1996vg}.  Although the conditions required for vascular-like development in these in vitro culture systems are well established, the mechanisms driving pattern formation of endothelial cells are heavily debated, and a wide range of plausible mechanisms has been proposed in the form of mathematical and computational models reproducing aspects of angiogenesis (reviewed in \cite{RMHMerks:2009hl,Boas:2012gg,Scianna:2013ho}).

Typical ingredients of network formation models are (a) an attractive force between endothelial cells, which is (b) proportional to the cell density, and (c) inhibited or attenuated at higher cellular densities. The attractive force can be due to mechanical traction or due to chemotaxis. Manoussaki, Murray, and coworkers \cite{Manoussaki:1996wm,Manoussaki:2003iy} proposed a mechanical model of angiogenic network formation, based on the Oster and Murray \cite{Oster:1983wf,Murray:1983wu} continuum mechanics theory of morphogenesis. In their model, endothelial cells  exert a uniform traction force on the ECM, dragging the ECM and the associated endothelial cells towards them. The traction forces saturated at a maximum cell density. Namy and coworkers\cite{Namy:2004im} replaced the endothelial cells' passive motion along with the ECM for active cell motility via haptotaxis, in which cells move actively towards higher concentrations of the ECM. Both models also included a strain-biased random walk term for the endothelial cells, but they found that it had little effect on network formation; the mechanism was dominated by cell aggregation. In their model based on chemotaxis, Preziosi and coworkers \cite{Gamba:2003dz,Serini:2003hq} assumed that cells attract one another via the secreted chemoattractant VEGF. Due to diffusion and first-order degradation, the chemoattractant forms exponential gradients around cells leading to cell aggregation in much the same way as that assumed in the Manoussaki and Namy models. These chemotaxis-based hypotheses formed the basis for a series of cell-based models based on the cellular Potts model (CPM). Assuming chemotactic cell-cell attraction, and a biologically-plausible overdamped cell motility, the cells in these CPM models form round aggregates,  in accordance with the  Keller-Segel model of cell aggregation \cite{Keller:1970vv}. Additional assumptions, including an elongated cell shape \cite{RMHMerks:2006jp} or contact inhibition of chemotaxis \cite{RMHMerks:2008fv} are needed to transform these circular aggregates into vascular-like network patterns. Related network formation models studied the role of ECM-bound growth factors \cite{KohnLuque:2011fm,KohnLuque:2013fw,Kleinstreuer:2013if} and a range of additional secreted and exogenous growth factors \cite{Kleinstreuer:2013if}, and studied the ability of the contact-inhibition mechanism to produce three-dimensional blood-vessel-like structures \cite{Singh:2013vh}. Szab{\'o} and coworkers found that in culture, astroglia-related rat C6 cells and muscle-related mouse C2C12 cells organize into network-like structures on rigid culture substrates \cite{Szabo:2007ib}, such that ECM-density or chemoattractant gradients are excluded. They proposed a model where cells were preferentially attracted to or preferentially adhered to locally elongated structures. As an alternative mechanism for ``gel-free'' network formation it was found that elongated cells can also produce networks in absence of chemoattractant gradients \cite{Palm:2013cy}. 

Paradoxically, despite the diverse assumptions underlying the mathematical models proposed for vascular network formation, many are at least partly supported by experimental evidence. This suggests that a combination of chemotaxis, and chemical and mechanical cell-ECM interactions drives network formation, or that each alternative mechanism operates in  a different tissue, developmental stage, or culture condition.
A problem is that one mathematical representation may represent a range of equivalent alternative underlying mechanisms. For example, a model representing cell-cell attraction cannot distinguish between chemotaxis-based cellular attraction \cite{Gamba:2003dz,Serini:2003hq,RMHMerks:2006jp,RMHMerks:2008fv}, attraction via haptotaxis \cite{Namy:2004im}, direct mechanical attraction \cite{JDMurray:1998tr,Manoussaki:1996wm} or cell shape dependent adhesion \cite{Szabo:2007ib,Szabo:2008em}, because the basic principles underlying these models are equivalent \cite{RMHMerks:2008fv,RMHMerks:2009hl}. 
As a solution to this problem, a sufficiently correct complete description of endothelial cell behavior should suffice for the emergence of the subsequent levels of organization of the system, an approach that requires that the system has been experimentally characterized at all levels of organization. 

The role of cell traction and ECM mechanics during in vitro angiogenesis have been characterized experimentally particularly well, making it a good starting point for such a multiscale approach. Endothelial cells apply traction forces on the extracellular matrix, as demonstrated by a variety of techniques, e.g., wrinkle formation on elastic substrates \cite{Vernon.ajp95}, force-generation on micropillar substrates \cite{2003PNAS..100.1484T}, and traction force microscopy \cite{ReinhartKing:2005dq,Califano:2008ct}. Using scanning electron microscopy, Vernon and Sage \cite{Vernon.ajp95} found that ECM ribbons radiate from endothelial cells cultured in Matrigel, suggesting that the traction forces locally reorient the extracellular matrix. The cellular traction forces produce local strains in the matrix, which can affect the motility of nearby cells \cite{ReinhartKing:2008cv}. Thus endothelial cells can both generate, and respond to local strains in the extracellular matrix, suggesting a feedback loop that may act as a means for mechanical cell-cell communication \cite{ReinhartKing:2008cv} and hence coordinate collective cell behavior. Here, we use a hybrid cellular Potts and finite element model to show that a set of  assumptions mimicking mechanical cell-cell communication via the ECM suffices to reproduce observed single cell behavior \cite{Califano:2010if,Winer:2011et}, pairwise cell interactions \cite{ReinhartKing:2008cv}, and collective cell behavior: network formation and sprouting.

\section*{Results}

\subsection*{Response of endothelial cells to static strains in ECM}

First we set out to capture, at a phenomenological level, the response of endothelial cells to static strains in the ECM in absence of cellular traction forces. When grown on statically, uniaxially stretched collagen-enriched scaffolds, murine embryonic heart endothelial (H5V) cells orient in the direction of strain, whereas cells grown on unstrained scaffolds orient in random directions \cite{vanderSchaft:2011cx}. Because the collagen fibers make the scaffold stiffen in the direction of strain, we hypothesized that the observed alignment of cells is due to durotaxis, the propensity of cells to migrate up gradients of substrate rigidity \cite{Lo:2000cj} and to spread on stiff substrates \cite{Califano:2010dp,Pelham:1997uc}.  In our model we assumed (a) {\em strain stiffening}: a strained ECM is stiffer along the strain orientation than perpendicular to it, such that (b) due to durotaxis the endothelial cells preferentially extend pseudopods along the strain orientation, along which the ECM is stiffest,  giving cells the most grip.  To keep the ECM mechanics simulations computationally tractable, we assumed an isotropic and linearly elastic ECM. With these assumptions it is not possible to model strain stiffening explicitly. We therefore mimicked strain stiffening by assuming that stiffness is an increasing, linear function of the local strain.  

Durotaxis was modelled as follows, to reflect the observation that focal adhesion maturation occurs under the influence of local tension \cite{Riveline:2001bp}: At low local stiffness, we applied standard cellular Potts dynamics to mimic the iterative formation and breakdown of ECM adhesions, producing ``fluctuating'' pseudopods. However, if the stiffness was enhanced locally, we assumed that the resulting tension in the pseudopod led to maturation of the focal adhesion \cite{Riveline:2001bp,Kuo:2011fb}, stabilizing the pseudopod as long as the tension persists. To mimic such focal adhesion maturation in the cellular Potts model, we increased the probability of extension along the local strain orientation, and reduced the probability of retraction (see Methods for detail).

\begin{figure}[!htbp]
\begin{center}
\includegraphics[width=\textwidth]{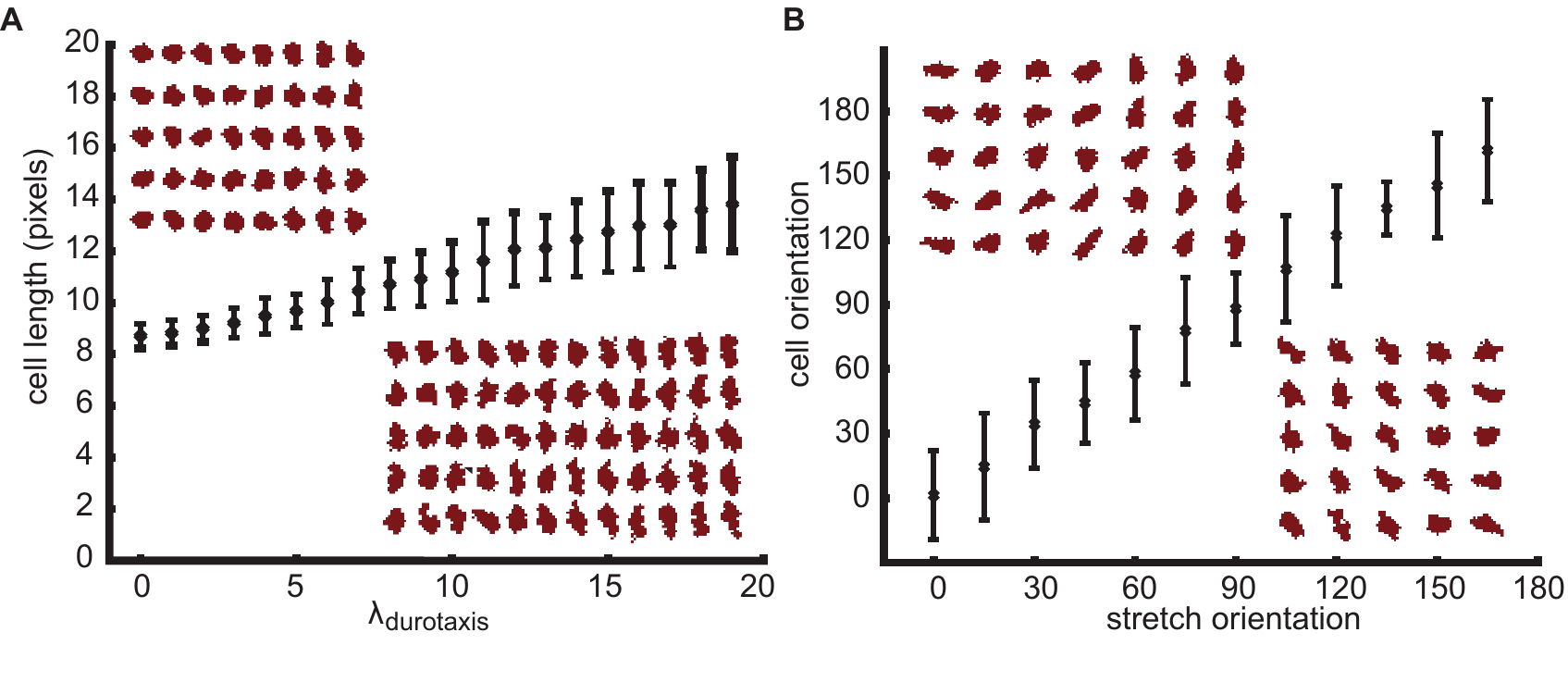}
\end{center}
\caption{{\bf Simulated cellular responses to static strains.} Cells do not generate traction forces in this figure. {\it (A)} Cell length as a function of the durotaxis parameter, $\lambda_{\mathrm{durotaxis}}$, on a substrate stretched along the vertical axis. {\it (B)} Cell orientation as a function of the stretch orientation (simulated with $\lambda_\mathrm{durotaxis}=10$). Error bars show standard deviation for $n=100$. Insets show five simulations per value tested.}
 \label{fig:static-stress}
 \end{figure}

Figure~\ref{fig:static-stress}~{\it A} shows the response of the simulated cells to uniaxial stretch along the vertical axis. With increasing values of the durotaxis parameter $\lambda_\mathrm{durotaxis}$ (see Eq.~\ref{eq:strain}), the endothelial cells elongate more. To test the sensitivity of the durotaxis model for lattice effects, we varied the orientation of the applied strain over a range $[0-180]^{\circ}$ and measured the resulting orientation of the cells.  Figure~\ref{fig:static-stress} shows that the average orientation of the cells follows the orientation of the stretch isotropically. Thus the durotaxis component of our model phenomenologically reproduces published responses of endothelial cells to uniaxial stretch \cite{vanderSchaft:2011cx}.

\subsection*{Generation of strains in ECM due to cellular traction}
We next attempted to mimic the forces applied by cells onto the extracellular matrix, in absence of durotaxis. Traction-force microscopy experiments \cite{ReinhartKing:2005dq,Califano:2010dp} show that endothelial cells contract and exert tensional forces on the ECM. The forces are typically directed inward, towards the center of the cell, and forces concentrate at the tips of pseudopods. A recent modeling study by Lemmon and Romer \cite{Lemmon:2010ju} found that an accurate prediction of the direction and relative magnitudes of these traction forces within the cell can be obtained by assuming that each lattice node $i$ covered by the cell pulls on every other node the cell covers, $j$, with a force proportional to their distance, $d_{i,j}$. Because this model gives experimentally plausible predictions for fibroblasts, endothelial cells, and keratocytes \cite{Lemmon:2010ju}, we adopted it to mimic the cell-shape dependent contractile forces that endothelial cells exert onto the ECM. Figure~\ref{fig:traction-forces} shows the contractile forces ({\it black}) and resulting ECM strains ({\it blue}) generated in our model by two adjacent cells. The traction forces and ECM strains become largest at the cellular ``pseudopods'', qualitatively agreeing with traction force fields reported for endothelial cells \cite{ReinhartKing:2005dq}.

\subsection*{Mechanical cell-ECM feedback qualitatively reproduces effect of substrate stiffness on cell shape and motility}

The two previous sections discussed how the simulated cells can respond to and induce strain in the ECM in an experimentally plausible way. To test how the simulated cells respond to the strains they generate themselves, we studied the behavior of simulated, single cells in presence of both the cell traction mechanisms and the durotaxis mechanisms. During each time step, we used the Lemmon and Romer \cite{Lemmon:2010ju} model to calculate traction forces corresponding to current cell positions. Next, we started the finite element analysis from an undeformed matrix, calculating steady-state strains for the current traction forces. To simulate cell movement, which was biased by the local matrix strains using the durotaxis mechanism, we then applied one cell motility simulation time step, or Monte Carlo step (MCS; the MCS is the unit of time of our simulation; see Methods for detail and Discussion for an estimate of the real time corresponding to an MCS). After running the CPM for one MCS we again relaxed the matrix such that the next step started with an undeformed matrix. Thus we currently did not consider cell memory of substrate strains.

As Figure~\ref{fig:single-cells} and Video S1 demonstrate, in this model matrix stiffness affects both the morphology and motility of the simulated cells. On the most compliant substrate tested (0.5 kPa) the simulated cells contract and round up, whereas cells spread isotropically on the stiffest substrate tested (32 kPa).   Overall, the cellular area increases with substrate stiffness (Figure~\ref{fig:single-cells}~{\it B}). On matrices of intermediate stiffnesses (around 12 kPa) the cells elongate, as reflected by measurements of the cell length (Figure~\ref{fig:single-cells}~{\it C}) and eccentricity (Figure~\ref{fig:single-cells}~{\it D}) that both have maximum values at around 12 kPa.  Such a biphasic dependence of cellular morphology on the stiffness of the ECM mimics the behavior of endothelial cells \cite{Califano:2010dp} and cardiac myocytes \cite{Winer:2011et} in matrices of varying stiffness. The dependence of cell shapes on  substrate stiffnesses is due to the transition from fluctuating to adherent pseudopods with increasing stiffness. Focal adhesions of cells on soft substrates all remain in the ``fluctuating'' state, irrespective of the local strains. On intermediate substrates, some pseudopods, due to increased traction, move to an extended state (mimicking a mature focal adhesion), generating more traction in this direction. Hence an initial stochastic elongation self-enhances in a feedback loop of increasing traction and strain stiffening. Such a self-enhancing cell-elongation starting from an initial anisotropy in cell spreading has previously been suggested by Winer et al \cite{Winer:2009jg}. Extensions perpendicular to the long axis of an elongated cell do not occur since there is insufficient traction and the volume constraint is limiting. 
At matrices of high stiffness all pseudopods attempt to extend, mimicking the formation of static focal adhesion, until the volume constraint becomes limiting.  This makes the cells spread more on stiff substrates than on soft substrates, with weaker volume constraints (lower values of $\lambda$) producing a stronger effect of substrate stiffness on cell shape and cell area (Figure S1) .

We also measured the random motility of the cells by characterizing their dispersion coefficients, which we derived from the mean square displacements of the cells (Figure S2; see section Morphometry for detail). The dispersion coefficients show biphasic behavior, with the highest motilities occurring at around 12 kPa (Figure~\ref{fig:single-cells}~{\it E}). The biphasic dependence of the dispersion to substrate stiffness is in accordance with in vitro behavior of neutrophils \cite{Stroka:2009jr}, and smooth muscle cells \cite{Peyton:2005jx}. Here it is typically thought to be due to a balance of adhesion and actin polymerization, or due to the interplay between focal adhesion dynamics and myosin-based contractility \cite{Stroka:2009jr}. In our model, the effect is more likely due to the appearance of eccentric cell shapes at intermediate stiffnesses; as a result, only the tips of the cell generate sufficient strain in the matrix to extend pseudopods, producing more persistent motion than the round cells at stiff or soft substrates. It will be interesting to see if a similar relationship between cell shape and cell motility holds in vitro.  Thus the model rules for cell traction and stretch guidance based on durotaxis and strain stiffening suffice to reproduce an experimentally plausible cellular response to matrix stiffness.

\subsection*{Mechanical cell-ECM feedback coordinates behavior of adjacent cells}
Strains induced by endothelial cells  on a compliant substrate with low concentrations of arginine-glycine-aspartic acid(RGD)-containing nonapeptides can affect the behavior of adjacent cells
\cite{ReinhartKing:2008cv}. On soft substrates (5.5 kPa or below) the cells reduced
the motility of adjacent cells, whereas on stiff substrates (33 kPa) such an
effect was not found. On substrates of intermediate stiffness  (5.5 kPa), 
adjacent endothelial cells repeatedly attached and
detached from one another, and cells moved more slowly in close vicinity of other cells, than when they were on their own. Because the extent to which cells could affect the motility of nearby cells depended on matrix compliancy,
mechanical traction forces could act as a means for cell-cell
communication \cite{ReinhartKing:2008cv}. To test if the simple strain-based mechanism represented in our model
suffices for reproducing such mechanical cell-cell communication, we
initiated the simulations with pairs of cells placed adjacent to one
another at a distance of fourteen lattice sites corresponding to a distance
of $35\;\mathrm{\mu m}$, and ran a series of simulations on substrates of
varying stiffness (Figure~\ref{fig:cellpairs}~{\it A} and Video S2). 

The cells behaved similar to the single cell simulations
(Figure~\ref{fig:single-cells}), with little cell-cell interactions at the
lower and higher stiffness ranges. Consistent with previous observations \cite{ReinhartKing:2008cv}, cell pairs on substrates of intermediate stiffness (12 kPa) dispersed more slowly than individual cells (paired two-sample {\it t}-test at 5000 MCS, $p<0.05$ for 12 kPa), whereas individual cells and cell pairs dispersed at indistinguishable ($p>0.05$) rates on stiff (14 kPa or more) or soft (10 kPa or below) substrates (Figure~\ref{fig:cellpairs},~{\it B-D}) and Figure~S3).

Also in agreement with the previous, experimental observations \cite{ReinhartKing:2008cv}, on
a simulated substrate of intermediate stiffness (12 kPa) the cells responded to
the matrix strains induced by the adjacent cell by repeatedly
touching each other, and separating again (Figure~\ref{fig:cellpairs}~{\it E}). The contact duration of cells on soft and stiff substrates, when they get close enough to each other, are typically longer than for intermediate substrates. This behavior is also similar to observations in vitro\cite{ReinhartKing:2008cv}. As one might expect that strongly adherent cells will not repeatedly touch and retract, but rather stay connected upon first contact, we investigated the effect of cell adhesion on these parameters (Figure~S4). Consistent with this intuition, for stronger adhesion, the contact count tends to be reduced and the contact durations tend to increase, but the overall trend holds: at intermediate matrix stiffnesses we continue to observe more frequent cell contacts than for more soft or more stiff matrices. Thus the observed pairwise cell behavior is primarily driven by durotaxis.

Mechanical strain can also coordinate the relative orientation of cells. Fibroblasts seeded on a compliant gel tend to align in a head-to-tail
fashion along the orientation of mechanical strain \cite{Takakuda:1996gl}. Bischofs and Schwarz \cite{Bischofs:636529} proposed a computational model to explain this observation. Their model assumes that cells prefer the direction of maximal effective stiffness, where the cell has to do the least work to build up a force. This work is minimal between two aligned cells, because maximum strain stiffening occurs along the axis of contraction.  Interestingly, visualization of our model results (Figure~\ref{fig:static-stress}~{\it C}) suggested similar head-to-tail
alignment of our model cells at around 12 kPa. To quantify cell alignment in our simulations, we measured the angle $\alpha$ between the lines $l_1$ and $l_2$, defining the long axes of the cells and crossing the centers of mass of the cells (Figure~\ref{fig:cellpairs}~{\it F}). We
classified the angles as acute ($\alpha < \pi/2$; {\em i.e.} no
alignment) or obtuse ($\alpha \geq \pi/2$; alignment). At matrix
stiffnesses up to around 10 kPa, about one fourth of the angles $\alpha$ were
obtuse, corresponding to the expected value for uncorrelated cell
orientations.  However, at 12 kPa and 14 kPa significantly more than
a fourth of the angles $\alpha$ between the cell axes were obtuse (55/100 for
12 kPa, $p<1\times10^{-8}$ and 52/100 for 14 kPa, $p<1\times10^{-8}$, binomial test), and for substrate compliancies of 8 to 16 kPa significantly more of the angles $\alpha$ were obtuse than for 4 kPa ($p<0.01$ for 8 kPa, and $p<1\times10^{-12}$ for 10 kPa to 16 kPa; two-tailed Welch's t-test), suggesting that the mechanical coupling represented in our model causes cells to align in a head-to-tail fashion. 

\subsection*{Mechanical cell-cell communication drives biologically-realistic collective cell behavior}

After observing that the local, mechanical cell-ECM interactions assumed in our model sufficed for correctly reproducing many aspects of the behavior of individual endothelial cells on compliant matrices and of the mechanical communication of pairs of endothelial cells on compliant matrices, we asked what collective cell behavior the mechanical cell-cell coordination produced. When seeded subconfluently onto a compliant matrix (e.g., Matrigel), endothelial cells tend to organize into polygonal, vascular-like networks \cite{Folkman:1980vf,Kubota:1988ve,Califano:2008ct,Parsa:2011ge}.  To mimic such endothelial cell cultures, we initialized our simulations with (approximately) 450 cells uniformly distributed over a lattice of $300\times300$ pixels ($0.75\times0.75\;\mathrm{mm^2}$), corresponding to a cell density of 800 endothelial cells per $\mathrm{mm^2}$. In accordance with experimental observations on gels with low concentrations of collagen \cite{Califano:2008ct} or RGD-peptides \cite{ReinhartKing:2008cv}, after 3000 MCS networks had not formed on soft matrices (0.5-4 kPa) or on stiff matrices (16-32 kPa) (Figure~\ref{fig:networks}~{\it A}): The cells tended to form small clusters (Figure~\ref{fig:networks}~{\it A}). Interestingly, on matrices of intermediate stiffness after around 300 MCS the cells organized into chains (8 kPa) or network-like structures (10 kPa and 12 kPa) similar to vascular network-like structures observed in endothelial cell cultures \cite{Folkman:1980vf,Kubota:1988ve,Califano:2008ct,Parsa:2011ge}. The optimal stiffness ($\approx \mathrm{10\;kPa}$) for network formation is slightly lower than the stiffness of the substrate ($\approx \mathrm{12\;kPa}$) on which single cells elongate the most (Figure~\ref{fig:single-cells}~{\it A}). In comparison with a single cell, the collective pulling of a cell colony creates larger strains in the substrate. Consequently, the strain threshold inducing cell elongation is crossed at smaller substrate stiffness.

Figure~\ref{fig:networks}~{\it B} and Video S3 show a time-lapse of the development of a network configuration on a substrate of 10kPa. The cells organized into a network structure within a few hundred MCS. The network was dynamically stable, with minor remodeling events taking place, including closure and bridging of lacunae. Figure~\ref{fig:networks}~{\it C} shows such a bridging event in detail.  In an existing lacuna (1800 MCS) stretch lines bridged the lacuna, and connected two groups of cells penetrating the lacuna (1980 MCS). The cells preferentially followed the path formed by these stretch lines (2150 MCS) and reached the other side of the lacuna by 2400 MCS. Such bridging events visually resemble sprouting in bovine endothelial cell cultures on compliant matrices (Figure~\ref{fig:networks}~{\it D}, Video S4, and~\cite{Califano:2008ct}). To stay close to the experimental conditions used for the observations of pairwise endothelial cell-cell interaction on compliant substrates \cite{ReinhartKing:2008cv} that we compared the simulations of pairwise interactions with, in this experiment we used a 2.5 kPa gel functionalized with $5\;\mathrm{\mu g/ml}$ RGD peptide  - a stiffness at which no network-formation is found in our simulations. Although we thus do not yet reach full quantitative agreement between model and experiment, note that network formation occurs at substrate stiffness of 10kPa on polyacrylamide matrices enriched with a low ($1\;\mathrm{\mu g/ml}$) concentration of collagen \cite{Califano:2008ct}. 

We next asked if the mechanical model could also reproduce sprouting from endothelial spheroids \cite{Korff:1999ty,Kniazeva:2009jm}. Video S5 and Figure~\ref{fig:sprouting} shows the results of simulations initiated with a two-dimensional spheroid of cells after 3000 MCS. On soft (0.5-8 kPa) and on stiff (32 kPa) matrices the spheroids stayed intact over the time course of the simulation. On matrices of intermediary stiffness (10-12 kPa) the spheroids formed distinct sprouts, visually resembling the formation of sprouts in in vitro endothelial spheroids \cite{Korff:1999ty,Kniazeva:2009jm}. On the 14 kPa and 16 kPa matrices the cells migrated away from the spheroid, with some cell alignment still visible for the 14 kPa matrices. Observation of a sprout protruding from a spheroid at 10 kPa  suggests that  a new sprout starts when one of the cells at the edge of the cluster protrudes and increases the strain in front of it. In a positive feedback loop via an increase in perceived stiffness the strain guides the protruding cell forward. The strain in its wake then guides the other cells along (Figure~\ref{fig:sprouting}~{\it C}).

\section*{Discussion}
In this paper we introduced a computational model of the in vitro collective behavior of endothelial cells seeded on compliant substrates. The model is based on the experimentally supported assumptions that (a) endothelial cells generate mechanical strains in the substrate \cite{Lemmon:2010ju,ReinhartKing:2005dq}, (b) they perceive a stiffening of the substate along the strain orientation, and (c) they extend preferentially on stiffer substrate \cite{vanderSchaft:2011cx}. Thus, in short, the assumptions are: cell traction, strain stiffening, and durotaxis. The model simulations showed that these assumptions suffice to reproduce, in silico, experimentally observed behavior of endothelial cells at three higher level spatial scales: the single cell level, cell pairs, and the collective behavior of endothelial cells.  In accordance with experimental observation \cite{Califano:2010dp,Winer:2011et}, the simulated cells spread out on stiff matrices, they contracted on soft matrices, and  elongated on matrices of intermediate stiffness (Figure~\ref{fig:single-cells}). The same assumptions also sufficed to reproduce experimentally observed pairwise cell-cell coordination. On matrices of intermediate stiffness, endothelial cells slowed down each other (Figure~\ref{fig:cellpairs}~{\it B}) and repeatedly touched and retracted from each other (Figure~\ref{fig:cellpairs}~{\it E} and Video S2), in agreement with in vitro observations of bovine aortic endothelial cells on acrylamide gels \cite{ReinhartKing:2008cv}.  Also, in agreement with experimental observations of fibroblasts on compliant substrates \cite{Takakuda:1996gl} and previous model studies \cite{Bischofs:636529} the cells  repositioned into an aligned, head-to-tail orientation (Figure~\ref{fig:cellpairs}~{\it F}). The model simulations further suggest that these pairwise cell-cell interactions suffice for vascular-like network formation in vitro (Figure~\ref{fig:networks}) and sprouting of endothelial spheroids (Figure~\ref{fig:sprouting}). 

The correlation between pairwise cell-cell interactions and collective cell behavior observed in our computational model parallels observations in vitro. Cells elongate due to positive feedback between stretch-guided extension and cell traction, as previously suggested by Winer et al. \cite{Winer:2009jg}. Elongated and spindle-shaped cells are considered indicative of future cell network assembly \cite{Califano:2008ct}. Our model suggests that the elongated cell shapes produce oriented strains in the matrix, via which cells sense one another at a distance. In this way new connections are continuously formed over ``strain bridges"  (see, e.g., Figure~\ref{fig:networks}~{\it C,D} and Video S4), while other cellular connections break, producing dynamically stable networks as illustrated in Video S3. Such dynamic network restructuring was also observed during early embryonic development of the quail embryo \cite{PaulARupp:2004dy} and in bovine aortic endothelial cell cultures  (Figure~\ref{fig:networks}~{\it D} and~\cite{Califano:2008ct}), but not in human umbilical vein endothelial cell cultures \cite{RMHMerks:2006jp,Parsa:2011ge}. Also in agreement with experimental results, the collective behavior predicted by our model strongly depends on substrate stiffness. The strongest interaction between cell pairs is found on substrates of intermediate stiffness, enabling network formation \cite{ReinhartKing:2008cv}, whereas network assembly does not occur on stiffer or on softer substrates\cite{Califano:2008ct}. 

These agreements with experimental results are encouraging, but our model also lacks a number of properties of in vitro angiogenesis that pinpoint key components still missing from our description.  We compared the simulation of pairwise cell-cell interactions with previous experiments conducted on polyacrylamide gels, functionalized with RGD ligands \cite{ReinhartKing:2008cv}, which have linear elastic behavior for small deformations \cite{2005Natur.435..191S,Boudou:2006tm,Rudnicki:2013vp}. Strain-stiffening of polyacrylamide gels has been reported for deformations over $2\;\mathrm{\mu m}$ \cite{Boudou:2008ki}. Thus with pixels in our model measuring $2.5 \;\mathrm{\mu m}\times2.5 \;\mathrm{\mu m}$, strain-stiffening seems a reasonable assumption. Nevertheless, a possible alternative interpretation of the cell pair simulations is that the increased tension generated in pseudopods pulling on the matrix leads to a higher probability of focal adhesion maturation\cite{Riveline:2001bp,Kuo:2011fb}. A further issue is that in our simulations, single cells dispersed somewhat more quickly on soft gels than on stiff gels (Figure~\ref{fig:single-cells}~{\it E} and Figure~S2). This model behavior contradicts experimental observations that endothelial cells move fastest on stiff substrates \cite{ReinhartKing:2008cv}. Another open issue concerns the time scales of our simulations. In the present paper time we use the Monte Carlo step as a (computational) unit of time. To estimate the actual time corresponding to 1 MCS, we  scale the single cell dispersion coefficients shown in Figure~\ref{fig:single-cells}~{\it E} to experimental dispersion coefficients of bovine endothelial cells on compliant substrates in vitro \cite{ReinhartKing:2008cv}. Reported dispersion coefficients of endothelial cells range from around $1\;\mathrm{\mu m^2/min}$ (on substrates of $500\;\mathrm{Pa}$) to around $10\;\mathrm{\mu m^2/min}$ (on substrates of $5500\;\mathrm{Pa}$) (as derived from the MSDs in Figure~3a,c in \cite{ReinhartKing:2008cv} and based on $\mathrm{MSD}(t)=4Dt$; {\em cf.} Eq.~\ref{eq:dispersion}). The dispersion coefficients of single cells in our simulations are in the range of $0.03-0.08\;\mathrm{\mu m^2/MCS}$ (Figure~\ref{fig:single-cells}), assuming pixels of $2.5\times2.5\;\mathrm{\mu m^2}$.  Thus, based on fitting of single cell dispersion rates, the estimated length of 1 MCS is 0.5 to 3 seconds. The typical time scale of a vascular network formation simulation is around 3000 MCS (Figure~\ref{fig:networks}), {\em i.e.}, $12.5\;\mathrm{min}$ to $2.5\;\mathrm{hr}$ for these time scale estimates. In experiments, network formation takes longer, around $24\;\mathrm{hr}$. Thus in our current model the time scales of cell dispersion and network formation do not match exactly. A possible reason of this discrepancy is the short persistent length of cell motility in standard cellular Potts models. To better match the time scales of single cells and collective cell behavior in our model, in our future work we will increase the persistence length of the endothelial cells by using the available cellular Potts methodology \cite{ArielBalter:2007tf,Scianna:2011ke,Vroomans:2012jd}, or model the subcellular mechanisms of cell motility in more detail, e.g. by including mean-field models of actin polymerization \cite{Maree:2012jf,Maree:2006ek}. A further open issue is the interaction between substrate mechanics and cell-substrate adhesivity. Although the model correctly predicts the absence of network formation on stiff substrates, it cannot yet explain the observation that reducing the substrate adhesivity of the endothelial cells rescues network formation on stiff substrates \cite{Califano:2008ct}. On compliant gels endothelial cells must secrete fibronectin to form stable networks, whereas fibronectin polymerization inhibitors elicit spindle-like cellular phenotypes associated with network formation on stiff matrices, under conditions where networks do not normally form \cite{Califano:2008ct}. To explain these observations, straightforward future extensions of the model will include a more detailed description of cell-substrate adhesion, combined with models of ECM secretion and proteolysis \cite{KohnLuque:2011fm,Boas:2012gg,Daub:2013tj,Kleinstreuer:2013if}. 

The current model also assumes a uniform density (i.e., the infinitesimal strain assumption) and thickness of the extracellular matrix, whereas under some culture conditions the endothelial cells have been reported to pull the extracellular matrix underneath them \cite{RBVernon:1992ta}, producing gradient in matrix density and/or thickness. To describe the role of viscous deformations of the extracellular matrix in morphogenesis, Oster and Murray \cite{Oster:1983wf,Murray:1983wu} developed a continuum mechanical model of pattern formation in mesenchymal tissues. Their model assumed (a) that cells exert contractile forces onto the surrounding extracellular matrix, that will (b) locally deform the ECM, resulting in passive displacements of cells along with the ECM, and (c) produce density gradients in the ECM along which cells move actively due to haptotaxis. These mechanisms together produce periodic cell density patterns. Manoussaki et al. \cite{Manoussaki:1996wm} and Namy et al.\cite{Namy:2004im} applied this work to investigate mechanical cell-ECM interactions during angiogenesis, and demonstrated that the mechanism can produce vascular-like network patterns. In their model they also included an anisotropic diffusion term to simulate preferential movement along the local strain-direction, but the term was neither necessary nor sufficient for network formation. This finding contradicts our model in which strain-induced sprouting is the driving force of network formation and sprouting. The two models represent the two extremes of network formation on visco-elastic matrices. Here, the Manoussaki et al. \cite{Manoussaki:1996wm} and Namy et al. \cite{Namy:2004im} models represent patterning on viscous matrices, in which cellular traction forces pull the matrix together while inducing little strain or stress. Our model would represent elastic materials, in which pulling forces induce local strains. Future extensions of the model will include matrix remodelling (e.g., by assuming a matrix thickness field) allowing us to study the full range of viscoelastic matrices.

Apart from these biological issues, we made several mathematical simplifications that we will improve upon in future models of cell-ECM interactions. In the current model, for mathematical simplicity, we assumed that after each Monte Carlo step the matrix was undeformed again. Thus we currently did not consider cell memory of substrate strains. Further developments of the model presented here will improve on this issue, because actin filament dynamics are typically influenced by the past evolution of substrate deformations, e.g., due to reorientation of matrix fibers \cite{RBVernon:1992ta}. For computational efficiency, we assumed linearly elastic materials and infinitesimal strain in the finite element simulations, and mimicked durotaxis via a perceived strain-stiffening (Eq.~\ref{eq:strainstiffening}) where cells perceive increased ECM stiffness due to local strain. In our ongoing work we are interfacing the open source package FEBio ({\tt http://febio.org}) with the cellular Potts package CompuCell3D ({\tt http://compucell3D.org}). This will allow us to run our model with any ECM material available to users of FEBio, including strain-stiffening materials. Using an actual strain stiffening material may lead to longer-range interactions between cells, because locally stiffer regions may channel the stress between the cells \cite{Rudnicki:2011we}. A further technical limitation of our model is that we currently only run two-dimensional simulations,  representing cells moving on top of a two-dimensional culture system. The ongoing interfacing of FEBio and CompuCell3D will pave the way for modeling cell-ECM interactions in three-dimensional tissue cultures. We also plan to model  fibrous extracellular matrix materials in more detail.

A quite puzzling aspect of vascular network formation and spheroid sprouting is that so many alternative, often equally plausible computational models can explain it (reviewed in \cite{RMHMerks:2009hl}). Including the present model, there are at least three alternative computational models based on mechanical cell-ECM interactions \cite{Manoussaki:1996wm,JDMurray:1998tr,Manoussaki:2003iy,Namy:2004im,Tranqui:2000vx}, a series of models assuming chemoattraction between endothelial cells \cite{Gamba:2003dz,Serini:2003hq,Merks:2004wma,RMHMerks:2006jp,RMHMerks:2008fv,Guidolin:2009cu} and extensions thereof \cite{Scianna:2012ju,KohnLuque:2011fm,Kleinstreuer:2013if}, and models explaining network formation in absence of chemical or mechanical fields \cite{Szabo:2007ib,Szabo:2008em,Palm:2013cy}. Each of the models explains one aspect of vascular network formation or a response to an experimental treatment that the other models cannot explain, e.g. the relation between spindle-shaped cell phenotypes and network formation \cite{RMHMerks:2006jp,Palm:2013cy}, the requirement of VE-cadherin signaling for network formation and sprouting \cite{RMHMerks:2008fv,Szabo:2007ib}, the binding and release of growth factors from the ECM \cite{KohnLuque:2011fm,KohnLuque:2013fw}, the role of mechanical ECM restructuring and haptotaxis \cite{Manoussaki:1996wm,JDMurray:1998tr,Namy:2004im}, the response of vascular networks to toxins \cite{Kleinstreuer:2013if}, or the role of intracellular $\mathrm{Ca^{2+}}$ signaling \cite{Scianna:2011ke}. Among these alternative models, we must now experimentally falsify incorrect mechanisms, and fine-tune and possibly combine the remaining models to arrive at a more complete understanding of the mechanisms of angiogenesis. To this end, we are currently quantitatively comparing the kinetics of patterns produced by chemotaxis-based, traction-based, and cell-elongation based models with the kinetics of in vitro networks \cite{RMHMerks:2006jp,Parsa:2011ge}.  The resulting, more complete model would likely contain aspects of each of the available computational models and  assist in explaining the conflicting results obtained from the available experimental systems, culture conditions, and in silico models of angiogenesis.

\section*{Methods}

\label{sec:methods}
To model the biomechanical interactions between endothelial cells and compliant matrices, we developed a hybrid of the cellular Potts model (CPM) \cite{Graner:1992ve,JAGlazier:1993uk} to represent the stochastic motility of the endothelial cells, and a mechanical model based on the finite element method (FEM) \cite{Davies:2011wr} of the compliant extracellular matrix. Related CPM-FEM models were proposed  for the simulation of load-induced bone remodeling \cite{vanOers:2008do,vanOers:2008hj}, and recently a related approach was proposed in a model study of cell alignment \cite{Checa:2014jv}. A documented simulation code is provided as part of the Supporting Information (Supporting Text S1 and Code S1) and a detailed list of parameter values is given in Table S1.

\subsection*{Cellular Potts model}

The CPM represents cells on a regular square lattice, with one biological cell covering a cluster of connected lattice sites. To mimic random cell motility, the CPM iteratively expands and retracts the boundaries of the cells, depending on the passive forces acting on them and on the active forces exerted by the cells themselves. These are summarized in a balance of forces, represented by the Hamiltonian,

\begin{equation}
H= \sum_{\sigma\in\mathrm{cells}}\lambda\left(\frac{a(\sigma)-A(\sigma)}{A(\sigma)}\right)^2+\sum_{(\vec{x},\vec{x}\prime)} J(\sigma(\vec{x}),\sigma(\vec{x}\prime))(1-\delta(\sigma(\vec{x}),\sigma(\vec{x}\prime))).
\end{equation}

\noindent The first term is an (approximate) volume constraint, with $a(\sigma)$ the actual volume of the cells,  $A(\sigma)$, a resting volume, and ${\lambda}$  an elasticity parameter that regulates the permitted fluctuation around the resting volume. In contrast with the original formulation of the CPM~\cite{Graner:1992ve},  the deviation of the cell from its target volume is taken relative to the target volume, by analogy with the (non-dimensional) engineering strain. Alternative, similar volume constraints can be chosen \cite{Scianna:2012ju}. We use a value $A(\sigma)=50$ for all cells; the medium does not have a volume constraint. The second term represents cell-cell and cell-medium adhesion, where $J(\sigma(\vec{x}),\sigma(\vec{x}\prime))\geq0$ is the contact cost between two neighboring pixels, and $\delta$, the Kronecker delta.  Throughout the manuscript we use neutral cell-cell adhesion settings; $J(\sigma(\vec{x}),\sigma(\vec{x}\prime))=2.5$ at cell-cell interfaces, and $J(\sigma(\vec{x}),0)=1.25$ at cell-medium interfaces, with $\sigma(\vec{x})>0$ and $\sigma(\vec{x}\prime)>0$. In other words, cells have no preference for adhering to other cells or the medium. For these neutral cell adhesion parameter settings, cells will still adhere weakly to one another (a remedy for this effect was proposed in~\cite{2012PhBio...9a6010S}). Additional terms in the Hamiltonian represent the cells' responses to ECM mechanics, and will be described in more detail below.

The CPM iteratively selects a random lattice site $\vec{x}\prime$ and attempts to copy its state, $\sigma({\vec{x}\prime})$, into a randomly selected adjacent lattice site $\vec{x}$. To reflect the physical, ``passive'' behavioral response of the cells to their environment, the copy step is always accepted if it decreases the Hamiltonian. To account for the active random motility of biological cells, we allow for energetically unfavorable cell moves, by accepting copies that increase the Hamiltonian with Boltzmann probability,

\begin{equation}
P(\Delta H)= 
\begin{cases}
1 & \textrm{if}\;\Delta H <0 \\
e^{-\Delta H/T} & \textrm{if}\;\Delta H \geq 0,
\end{cases}
\end{equation}

\noindent where ${\Delta H}$ is the change in H if the copying were to occur, and $T>0$ parameterizes the intrinsic cell motility. It represents the extent to which the active cell motility can overcome the reactive forces (e.g. volume constraint or adhesions) in the environment. We assume that all cells keep the same motility and thus set $T$ to be constant throughout the simulations. During one Monte Carlo step (MCS), we perform $n$  copy attempts, with $n$ equal to the number of sites in the lattice. To prevent cells from splitting up into two or more disconnected patches, we use a connectivity constraint that rejects a spin flip $\sigma(\vec{x}\prime)\to \vec{x}$ if it would break apart the retracting cell $\sigma(\vec{x})$. 

\subsection*{Model of Compliant Substrate based on Finite Element Method}

A two-dimensional model describes the compliant substrate on which the cells move. Deformations are calculated using the finite element method (FEM; reviewed in \cite{Davies:2011wr}). The FEM represents the substrate as a lattice of finite elements, $e$, with each element corresponding to a pixel of the CPM. To obtain the finite element equations, the weak formulation (associated with the total potential energy) of the governing equations of the displacement $u$ of the substrate is set up, in order to obtain the finite element equations,

\begin{equation}
\dund{K} \uli{u}=\uli{f}, \label{eq:Kuf}
\end{equation}
 
\noindent with stiffness matrix $\dund{K}$, displacement $\uli{u}$, and forces $\uli{f}$. The vector $\uli{u} = [ u_{x_1}, u_{y_1}, u_{x_2}, ... u_{x_n}, u_{y_n} ] ^T$ contains the displacements of all nodes, which are the unknowns that the FEM calculates based on the active forces exerted onto the material, presented in $\uli{f}$. In this paper $\uli{f}$ only consists of traction forces that the cells apply onto the ECM, unless stated otherwise. In a two-dimensional analysis the forces $\uli{f}$ are divided by the thickness they are working on. For this we assume an effective substrate thickness $t=10\;\mathrm{\mu m}$. We impose boundary conditions of $\uli{u}=\uli{0}$ at the boundary of the CPM grid, this means that the substrate is fixed along the boundaries.

To a first approximation, in this work we consider an isotropic, uniform, linearly elastic substrate \cite{Ambrosi:2006tv,Bischofs:636529} and we apply infinitesimal strain theory: We assume that material properties, including local density and stiffness are unchanged by deformations. The global stiffness matrix $\dund{K}$ is assembled from the element stiffness matrices $\dund{K}_e$ (see Supporting Text S1 and~\cite{Davies:2011wr}), which describe the relation between nodes of each element, $e$,

 \begin{equation}
\dund{K}_e=\int_{\Omega_e}\dund{B}^T\dund{D}\dund{B} d\Omega_e.
 \end{equation}

\noindent where $\dund{B}$---the conventional strain-displacement matrix for a four-noded quadrilateral element (see Supporting Text S1 and~\cite{Davies:2011wr})---relates the node displacements $\uli{u}_e$ to the local strains, as,

\begin{equation}
 \uli{\epsilon}=\dund{B}\uli{u}_e.
 \label{eq:localstrains}
 \end{equation}
 
 \noindent The strain vector $\uli{\epsilon}$ is a column notation of the strain tensor $\dund{\epsilon}$ and $\dund{D}$ is the material property matrix. Assuming plane stress conditions,

\begin{equation}
\dund{D}=\frac{E}{1-v^2}
\begin{pmatrix} 1 & \nu & 0 \\ \nu & 1 & 0 \\ 0 & 0 & \frac{1}{2}(1-\nu),\end{pmatrix}
\end{equation}

\noindent where $E$ is the material's Young's modulus, and ${\nu}$ is Poisson's ratio. Throughout this study, we use a Poisson's ratio $\nu=0.45$ and Young's moduli ranging from $E=0.5\;\mathrm{kPa}$ to $E=32\;\mathrm{kPa}$, which are plausible values for most cell culture substrates \cite{Soofi:2009ik,Boudou:2006tm,Bischofs:636529}.  For more details of the derivation of Eq.~\ref{eq:Kuf}, and the entries in $\dund{B}$, see Supporting Text S1 and \cite{Davies:2011wr}. 

As a reference configuration for the displacements we used an unstretched substrate, $\uli{u}=\uli{0}$. Thus, after each Monte Carlo step (during which the cells positions and shapes have changed) the substrate is assumed to be undeformed, such that the stiffness matrix, $\dund{K}$, is constant in time. This prevents expensive calculations that would be necessary for recalculating $\dund{K}$ in each iteration. Although the previous displacements do not influence the new deformation of the substrate,  they are used as an initial guess for solving $\dund{K} \uli{u}=\uli{f}$, in order to reduce the number of iterations necessary to converge to the FEM solution.

\subsection*{Mechanical cell-substrate coupling}

\begin{figure}[!htbp]
\begin{center}
\includegraphics{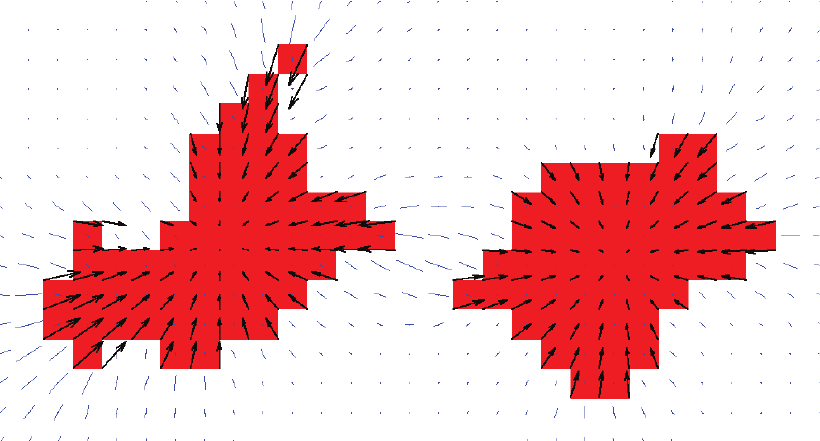}
\end{center}
\caption{{\bf Visualization of simulated traction forces ({\it black arrows}) and resulting matrix strains ({\it blue line segments}) generated in the proposed hybrid cellular Potts and finite element simulation model.}}
\label{fig:traction-forces}
\end{figure}

\begin{figure}[!htbp]
\begin{center}
\includegraphics[width=\textwidth]{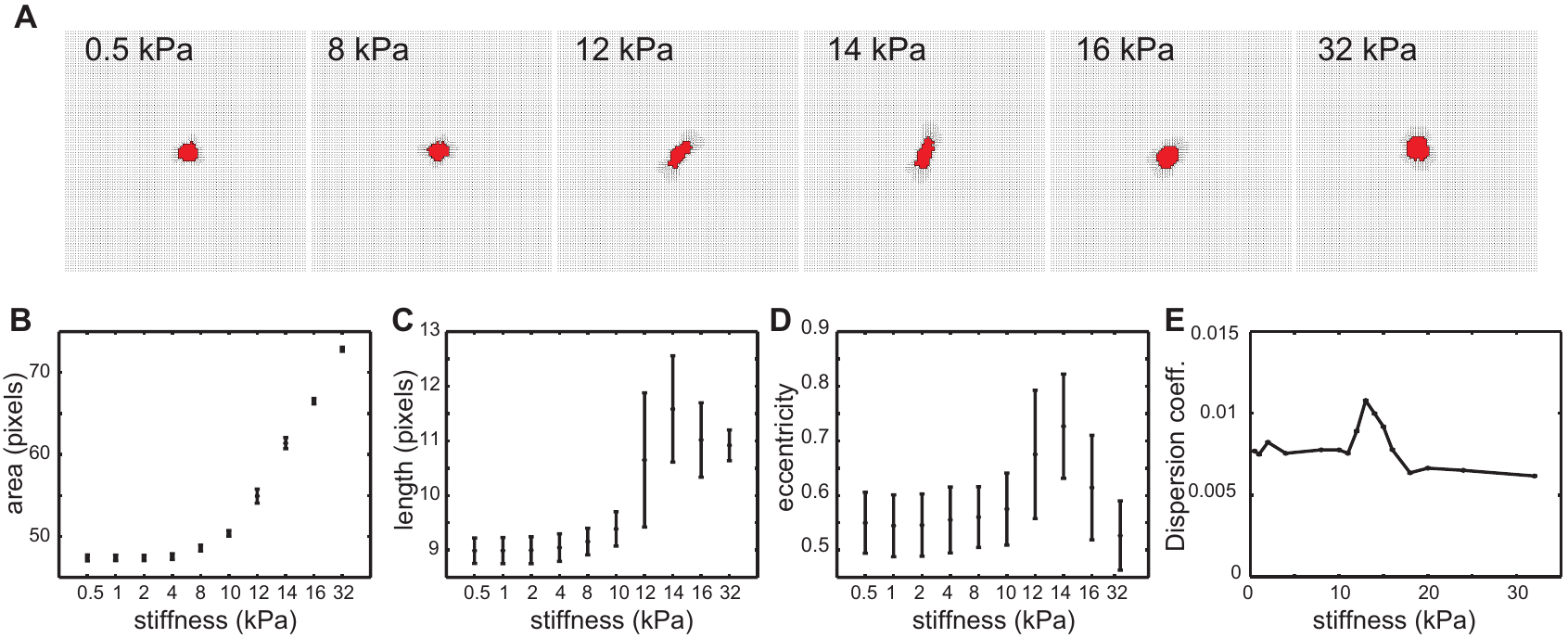}
\end{center}
\caption{{\bf Simulated individual cell responses to mechanical cell-ECM feedback.} {\it (A)} Single cells on substrates of varying stiffness after 100 MCS. Line pieces indicate strain magnitude and orientation. {\it (B)} cell area ($a(\sigma)$) of cells; {\it (C)} cell length (length of major axis if the cell is seen as an ellipse) as a function of substrate stiffness {\it (D)} cell eccentricity ($\xi=\sqrt{1-b^2/a^2}$, with $a$ and $b$ the lengths of the cell's major and minor semi-axes) as a function of stiffness. Mean and standard deviation shown for $n=100$ in panels B-D. {\it (E)} Dispersion coefficients of individual, simulated cells, derived from a linear fit on the mean square displacements (Figure S2); $n=1000$. Error bars indicate 95\% confidence intervals of linear fits.}
 \label{fig:single-cells}
\end{figure}

\begin{figure}[!htbp]
\begin{center}
\includegraphics[width=\textwidth]{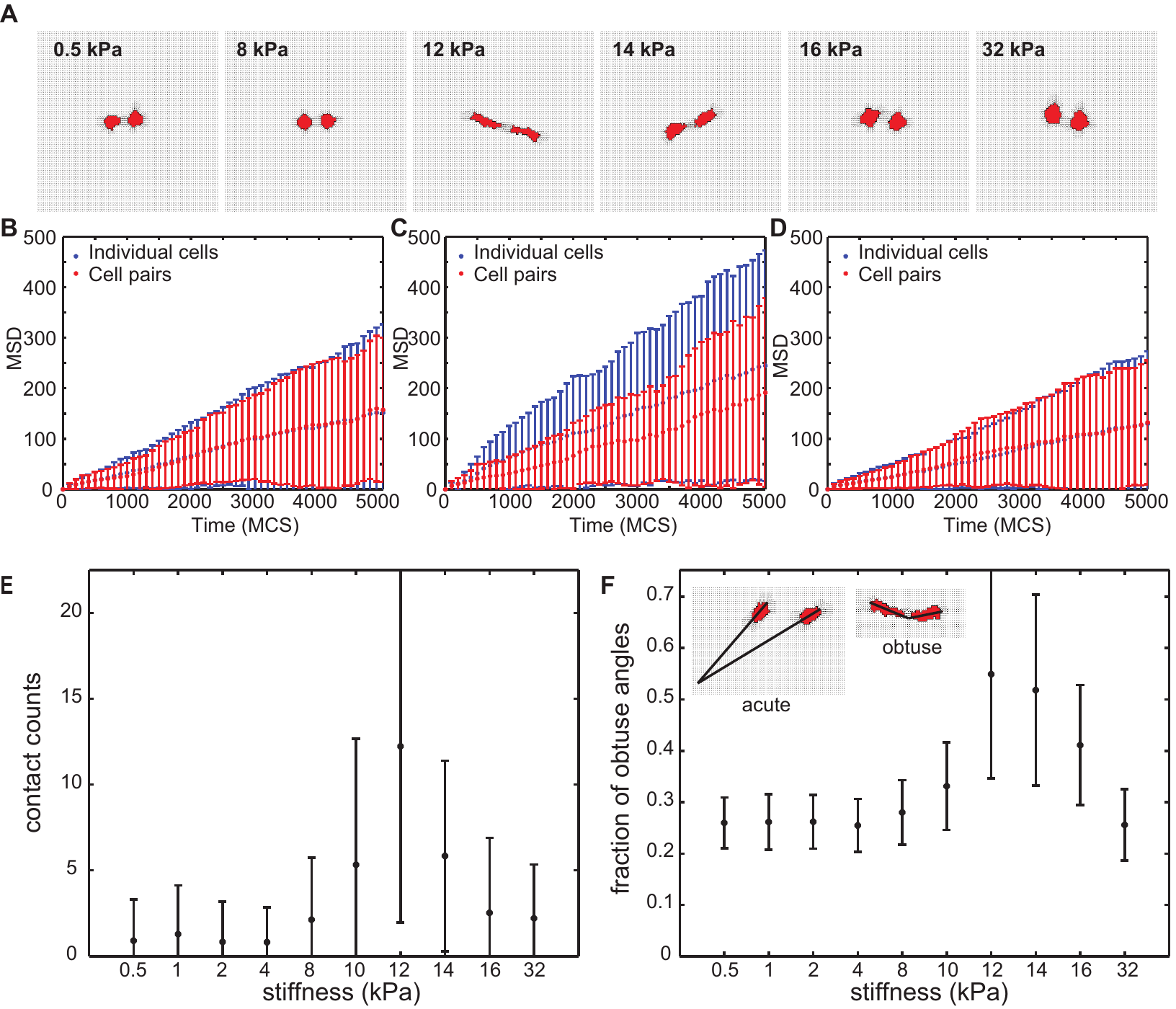}
 \end{center}
 \caption{{\bf Simulated cell-cell interactions on substrates of varying stiffnesses.} {\it (A)} Visualization of cell shapes and substrate strains in absence of external strain. Line pieces indicate strain magnitude and orientation. {\it (B-D)} Mean square displacement of individual cells ({\it blue errorbars}) and cell pairs ({\it red errorbars}) on simulated substrates. {\it (B)} 4 kPa; {\it (C)} 12 kPa; {\it (D)} 32 kPa. Error bars in panels B to D indicate standard deviation for $n=100$. {\it (E)} Number of cell-cell contacts made over 500 MCS between two simulated cells initiated at a distance of fourteen lattice sites from each other. Error bars show standard deviation over $n=100$ simulations {\it (F)} Quantification of head-to-tail alignment of cells. An obtuse angle between the two cells' long axes indicates that cells are oriented head-to-tail. Plotted is the fraction of Monte Carlo steps over MCS 20-500 that the two cells are aligned head-to-tail. Shown are means and standard deviations over 100 independent simulations on a field of 0.25 $\times$ 0.25 $\mathrm{mm^2}$ (100 $\times$ 100 pixels). Insets: examples of acute (left) and obtuse (right) cell configurations. }
  \label{fig:cellpairs}
 \end{figure}

To simulate cell-substrate feedback we alternate the cellular Potts model (CPM) steps with a simulation of the substrate deformations using the finite element method. We assume that cells apply a cell-shape dependent traction on the ECM and the cells respond to the resulting ECM strains by adjusting their cell shape. Using the CPM grid as the finite element mesh, the pixels of the CPM become four-node square elements in the FE-mesh. Adopting the model by Lemmon \& Romer \cite{Lemmon:2010ju}, we assume that each node $i$ covered by a CPM cell pulls on all other nodes $j$ in the same cell, at a force proportional to distance $\vec{d}_{i,j}$. The resultant force $\vec{F}_i$ on node $i$ then becomes,

\begin{equation}
\vec{F}_i=\mu\sum_j\vec{d}_{i,j},
\end{equation}

\noindent where $\Delta x$ is the lattice spacing and  ${\mu}$ gives the tension per unit length. This parameter has been scaled to $\mu=0.01\;\mathrm{nN/\mu m}$, such that the total cell traction corresponds to experimentally reported values \cite{AratynSchaus:2011gv}. 
The resultant forces point towards the cell centroid, and are proportional to the distance from it (Figure~\ref{fig:traction-forces}). In this way a CPM configuration yields a traction force $\uli{F}$, which are collected in the forces $\uli{f}$ for the finite element calculation. To calculate the resulting ECM strains, we solve $\dund{K}\uli{u}=\uli{f}$ for the node displacements $\uli{u}$ with a preconditioned conjugate gradient (PCG) solver \cite{Strang:1986uw}, and derive the local strains using Eq.~\ref{eq:localstrains}. The reference configuration for the displacements is an unstretched substrate, $\uli{u}=\uli{0}$. After a sufficiently accurate solution for the FEM equations has been obtained by the PCG solver, we run a Monte Carlo step of the CPM. After each MCS, which changes cell positions, the substrate is assumed to be undeformed again, for the sake of simplicity. Thus, the stiffness matrix, $\dund{K}$, is constant in time. 

We assume durotaxis, i.e., the CPM cells preferentially extend pseudopods on matrices of higher stiffness (e.g., because of strain stiffening). By analogy with chemotaxis algorithms \cite{Savill:1997vi} at the time of copying we add  the following durotaxis term to ${\Delta}H$ in response to the strain- and orientation-dependent ECM stiffness $E$,

\begin{equation}
\Delta H_{\mathrm{durotaxis}}= - g(\vec{x},\vec{x}\prime) \lambda_\mathrm{durotaxis}\left(h(E(\epsilon_1))(\vec{v}_1 \cdot \vec{v}_m)^2+h(E(\epsilon_2))(\vec{v}_2\cdot \vec{v}_m)^2\right),
\label{eq:strain}
\end{equation}

\noindent with $g(\vec{x},\vec{x}\prime)=1$ for extensions and $g(\vec{x},\vec{x}\prime)=-1$ for retractions,  $\lambda_\mathrm{durotaxis}$ is a parameter, $\vec{v}_m=\widehat{\vec{x}-\vec{x}\prime}$, a unit vector giving the copy direction, and $\epsilon_1$ and $\epsilon_2$, and $v_1$ and $v_2$ eigenvalues and eigenvectors of $\uli{\epsilon}$ representing the principal strains and strain orientation. We use the strain $\uli{\epsilon}(\vec{x})$ in the target pixel when considering an extension, and for retractions we use the strain in the source pixel, $\uli{\epsilon}(\vec{x}\prime)$. Thus we consider the strain in the ECM adjacent to the pseudopod. The sigmoid $h(E)= 1/(1+\exp (-\beta(E-E_\theta)))$, with threshold stiffness  $E_\theta$, and ${\beta}$, the steepness of the sigmoid, mimics maturation of focal adhesions under the influence of tension \cite{Riveline:2001bp}. The tension in focal adhesions will increase with higher local matrix stiffness, $E$, because the matrix will deform less easily.  The sigmoid function starts at zero, goes up when there is sufficient stiffness, and eventually reaches a maximum. This means that a certain level of stiffness is needed to cause a cell to spread. Alternative forms of $h(E)$ can be used: For an overview see Figure~S5. Due to limitations of our current finite element code and for reasons of computational efficiency, we assumed a linearly elastic, isotropic material in the FEM, thus precluding explicit strain stiffening effects in the FEM calculations. Instead, we implemented the effect of strain-stiffening in the cell response, where cells perceive increased ECM stiffness as a function of the principal strains $\epsilon_1$ and $\epsilon_2$,

\begin{equation}
E(\epsilon) = E_0 ( 1 + (\epsilon /  \epsilon_{st} )1_{\epsilon\geq0} )
\label{eq:strainstiffening}
\end{equation}

\noindent where $E_0$ sets a base stiffness for the substrate, and $\epsilon_\mathrm{st}$ is a stiffening parameter. The indicator function $1_{\epsilon>0}=\left\{1, \epsilon>0; 0, \epsilon\leq0\right\}$ indicates that strain stiffening of the substrate only occurs for substrate extensions ($\epsilon\geq0$); compression ($\epsilon<0$) does not stiffen or soften the substrate. 

\subsection*{Morphometry}

To characterize the random motility of single cells and cell pairs, we measured the cells' mean square displacement, 

\begin{equation}
\mathrm{MSD}(t)=\langle (\overline{C}(S,t)-\overline{C}(S,0))^2\rangle, 
\end{equation}

\noindent with $\overline{C}(S,t)$, the centroid of cell $S$ at Monte Carlo step ("time") $t$, given by

\begin{equation}
\overline{C}(S,t)=\frac{1}{|C(S,t)|}{\sum_{\vec{x}\in C(S,t)}\vec{x}},
\end{equation}

\noindent with $C(S,t)$, the set of coordinates of the lattice sites comprising cell $S$ at MCS $t$,

\begin{equation}
C(S,t)=\left\{\vec{x} : \vec{x}\in\mathbb{Z}^2 \land \sigma(\vec{x},t)=S\right\},
\end{equation}

\noindent and $\vec{x}=\left\{x_1,x_2\right\}$.  The MSD is a reliable measure of random motility \cite{Stokes:1991up} and it can be directly compared with experimental data (e.g.,~\cite{ReinhartKing:2008cv}). 

The dispersion coefficient, defined as 
\begin{equation}
D = \lim_{t \to \infty} \frac{1}{4t} \langle (\overline{C}(S,t) - \overline{C}(S,0))^2 \rangle,
\label{eq:dispersion}
\end{equation}
is derived from the slope of the MSD, and is used as a measure of the motility of random walkers. The length, orientation and eccentricity of cells were estimated from the inertia tensors $I(S)$ of the cells, defined as \cite{Zajac:2003bc},

\begin{equation}
I(S)=\left(
\begin{matrix}
\sum_{\vec{x}\in C(S)}(x_2-\overline{C}_2(S))^2 & -\sum_{\vec{x}\in C(S)}(x_1-\overline{C}_1(S))(x_2-\overline{C}_2(S)) \\
-\sum_{\vec{x}\in C(S)}(x_1-\overline{C}_1(S))(x_2-\overline{C}_2(S)) & \sum_{\vec{x}\in C(S)}(x_1-\overline{C}_1(S))^2
\end{matrix}\right).
\end{equation}

\noindent  Assuming cells are approximately ellipse-shaped, the length of cell $\sigma$ is approximated as $l(\sigma)=4 \sqrt{e_2(I(S))/|C(S)|}$, with $e_2(I(\sigma))$ the largest eigenvalue of $I(S)$. The eccentricity of a cell is defined using the eigenvalues of the inertia tensor $I(\sigma)$ as $\xi(\sigma) = \sqrt{1-\left({\frac{e_1(I(S))}{e_2(I(S))}}\right)^2}$, where $e_1(I(S))\leq e_2(I(S))$ are the eigenvalues of $I(S)$. An eccentricity close to zero corresponds to roughly circular cells and cells with an eccentricity close to unity are more elongated. The orientation of the cell is given by the eigenvectors of the inertia tensor $I(S)$.

\section*{Endothelial Cell Culture}
Bovine aortic endothelial cells (BAECs) (VEC Technologies, Rensselaer,
NY) were cultured through passage 12. Cells were kept at $37^{\circ}\mathrm{C}$ and 5\% $\mathrm{CO_2}$ and fed every other
day with Medium 199 (Invitrogen, Carlsbad, CA) supplemented with 10\% Fetal Clone III (HyClone, Logan, UT), 1\% MEM amino acids (Invitrogen), 1\% MEM vitamins (Medtech, Manassas, VA), and 1\% penicillin-streptomyocin (Invitrogen). Polyacrylamide hydrogels were synthesized as previously described \cite{Califano:2008ct}. Briefly, a gel mixture was prepared from MilliQ water, HEPES,
TEMED (Bio-Rad, Hercules, CA) and a 5\%:0.1\% ratio of acrylamide to
bis-acrylamide (Bio-Rad) to generate substrates with a Young's modulus
of 2,500 Pascals. Polymerization was initiated by the addition of
N-6-((acryloyl)amido)hexanoic acid (synthesized according to Pless et
al.~\cite{Pless:1983ws}) and ammonium persulfate (Bio-Rad) into the gel mixture.
Following polymerization, gels were incubated with $5\;\mathrm{\mu g/ml}$ RGD peptide
(GCGYGRGDSPG) (Genscript), followed by ethanolamine (Sigma). Gels were
stored in PBS overnight. Hydrogels were sterilized with ultraviolet light before cell culture. A
T-75 flask with a confluent BAEC monolayer was seeded onto the hydrogels
at 350,000 cells per gel (approximately 1,375 cells per
mm\textsuperscript{2}). The gels were maintained at $37^{\circ}\mathrm{C}$ and 5\% $\mathrm{CO_2}$ for three days prior
to imaging. ÊAfter replenishing with fresh complete media, the cells on hydrogels
were visualized with a Zeiss Axio Observer.Z1 inverted spinning disc
microscope with a Hamamatsu ORCA-R\textsuperscript{2} digital camera.
Images were captured every 30 minutes for 24 hours.

\section*{Acknowledgments}
The authors thank Sonja Boas for critical reading of the manuscript.
The investigations were supported by the Division for Earth and Life Sciences (ALW) with financial aid from the Netherlands Organization for Scientific Research (NWO), and by the NIH (grant number HL097296). This work was cofinanced by the Netherlands Consortium for Systems Biology (NCSB) which is part of the Netherlands Genomics Initiative / Netherlands organization for Scientific Research. 



\clearpage
\section*{Figures}

\begin{figure}[!htbp]
\begin{center}
\includegraphics[width=0.9\textwidth]{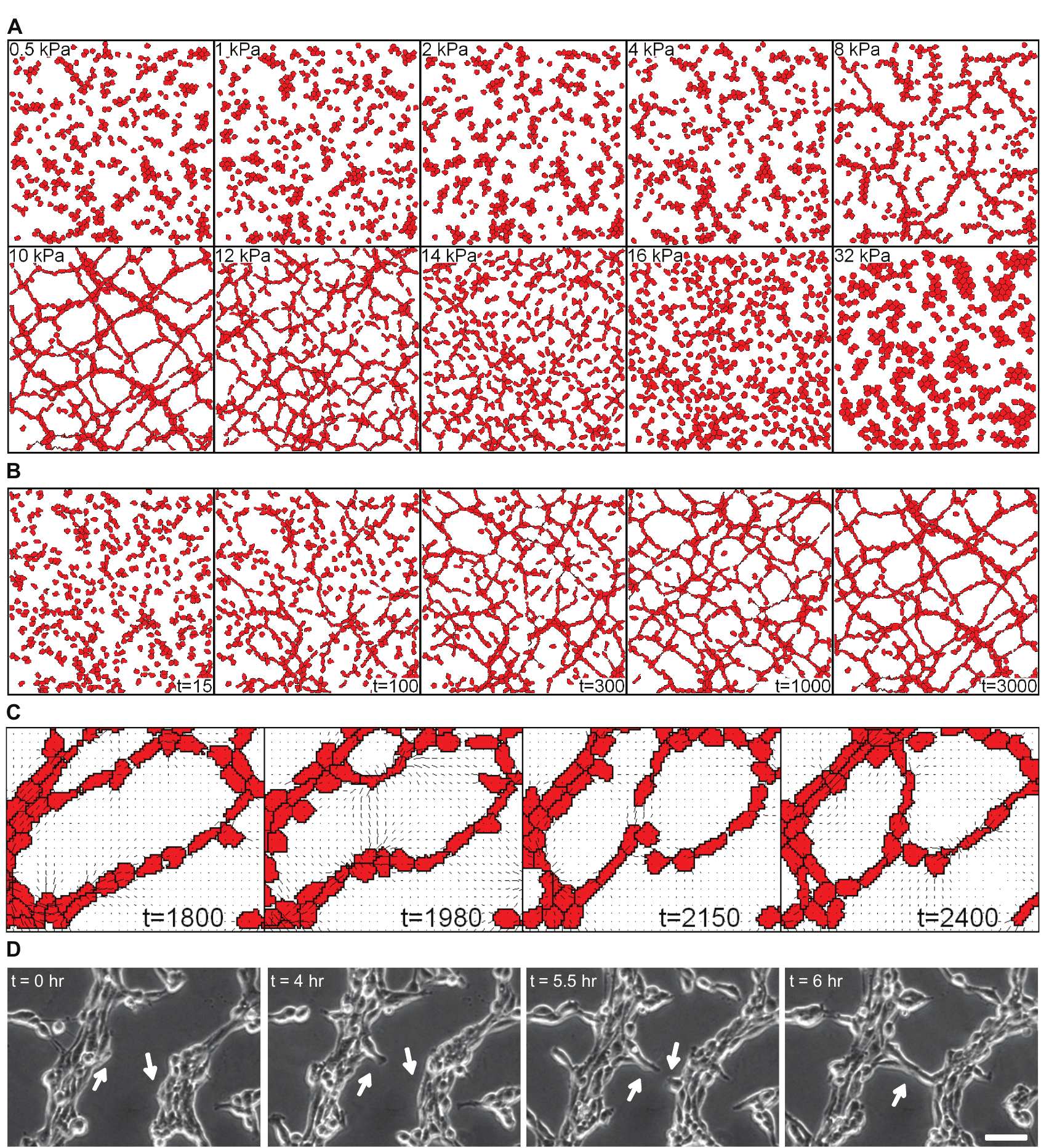}
\end{center}
\caption{{\bf Simulated network formation assay.} {\it (A)} Simulated collective cell behavior on substrates of varying stiffness, with a uniformly distributed initiated configuration of cells. {\it (B)} Time lapse showing the development of a polygonal network on a 10kPa substrate (time in MCS). Panels {\it A} and {\it B} represent a 0.75 $\times$ 0.75 $\mathrm{mm^2}$ area ($300\times300$ pixels) initiated with 450 cells. {\it (C)} Close-up of simulated network formation on a 10 kPa substrate, showing the reconnection of two sprouts. Time in MCS. {\it (D)} Time lapse imaging of bovine aortic endothelial cells seeded onto a 2.5 kPa polyacrylamide gel functionalized with RGD-peptide. Arrows indicate cells that join together and elongate into a network. Time scale is in hours. Scale bar is 50 $\mathrm{\mu m}$.}
\label{fig:networks}
\end{figure}

\begin{figure}[!htbp]
\begin{center}
\includegraphics[width=\textwidth]{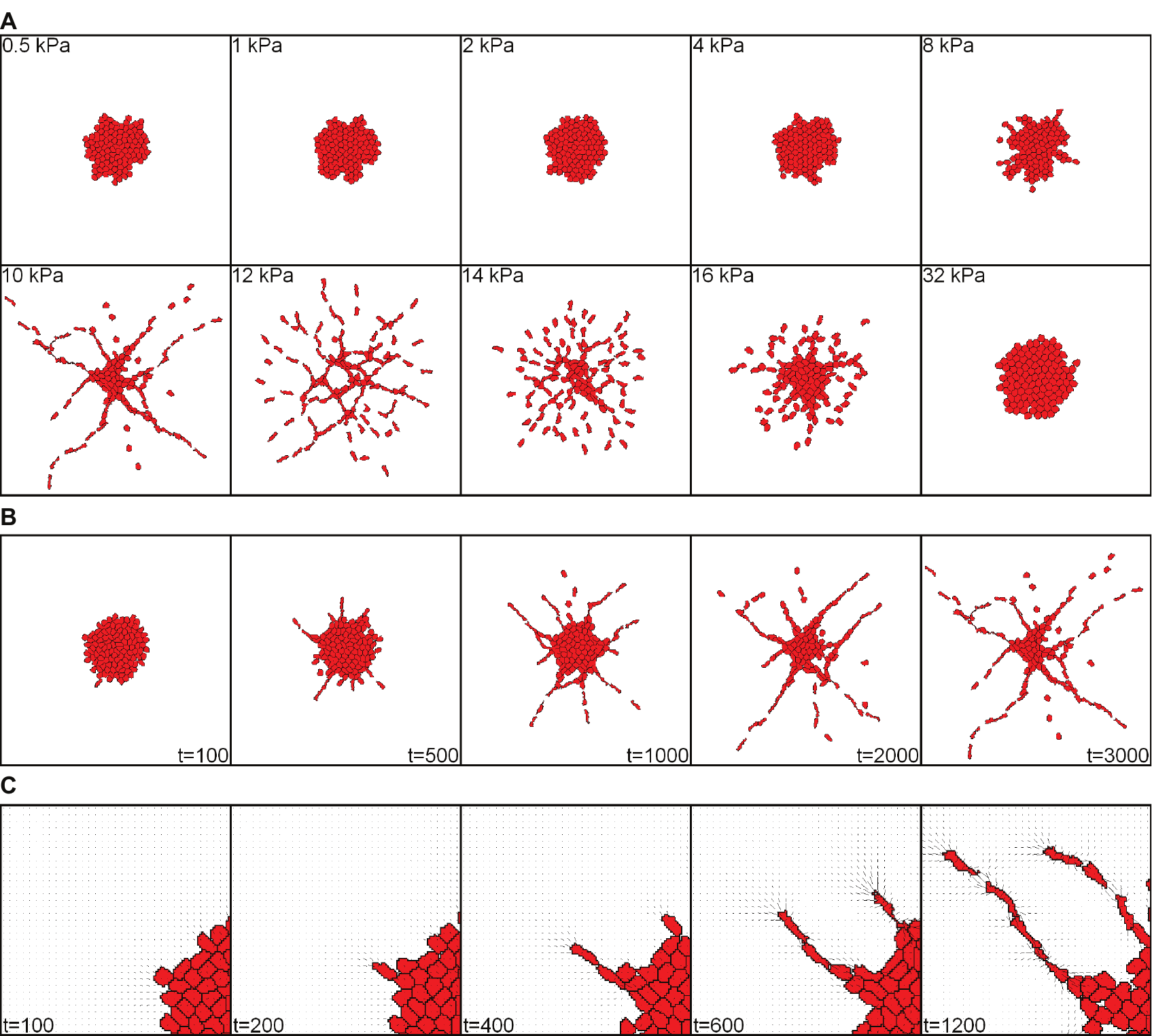}
\end{center}
\caption{{\bf Simulated spheroid assay.} {\it (A)} Collective behavior in a simulation initiated with a two-dimensional ``spheroid'' of cells, on substrates of varying stiffness. {\it (B)} Time lapse showing a sprouting spheroid on a 10kPa substrate. Time in MCS. Panels {\it A} and {\it B} represent a 0.75 $\times$ 0.75 $\mathrm{mm^2}$ area ($300\times300$ pixels) initiated with a spheroid consisting of 113 cells; {\it (C)} Close-up of sprouting on a 10 kPa substrate. Time in MCS. Black line pieces indicate strain magnitude and orientation.}
\label{fig:sprouting}
\end{figure}

\clearpage\section*{Supporting Information}

\noindent{\bf Table S1.} Parameter settings of the simulation model.
\bigskip

\noindent{\bf Figure S1.} Simulated responses of  individual cells to mechanical cell-ECM feedback as a function of the values of the volume restriction, $\lambda$. Columns:  area (left), cell length (middle) and eccentricity (right).  Mean and standard deviation shown for $n=100$ after 500 MCS on simulated substrates of stiffness varying from 0.5 kPa to 32 kPa. \bigskip

\noindent{\bf Figure S2.} Mean square displacements of individual cells on simulated substrates of stiffness varying from 0.5 kPa to 32 kPa.  Mean square displacement shown over $n=1000$ cells. \bigskip

\noindent{\bf Figure S3.} Mean square displacement of individual cells (blue errorbars) and cell pairs (red errorbars) on simulated substrates of stiffness varying from 0.5 kPa to 32 kPa. Error bars indicate standard deviation for $n=100$.
\bigskip

\noindent{\bf Figure S4.} Number of cell-cell contacts made over 500 MCS (left column) and contact duration (right column) over 500 MCS between two simulated cells initiated at a distance of fourteen lattice sites from each other on simulated substrates of stiffness varying from 0.5 kPa to 32 kPa, for intercellular contact energies varying from $J(\sigma(\vec{x}),\sigma(\vec{x}\prime))=0.5$ (adhesive cells) to $J(\sigma(\vec{x}),\sigma(\vec{x}\prime))=4$ (repulsive cells), with $\sigma(\vec{x})\geq1$ and $\sigma(\vec{x}\prime)\geq1$; $J(\sigma(\vec{x}),0)=1.25$ for all simulations. \bigskip

\noindent{\bf Figure S5.} Effect of form of model function $h(E)$ on cell shapes on substrates of different stiffnesses. {\it (A)} Standard, sigmoid function, as used in main text, $h(E)=1/(1+\exp(-\beta(E-E_0)))$ with $E_0=15000$, $\beta=0.0005$, and $\lambda_\mathrm{durotaxis}=10$. {\it (B)} Saturated function, $h(E) = (E/E_0)/(1+E/E_0)$, with $E_0=15000$ and $\lambda_\mathrm{durotaxis}=25$. {\it (C)} Piecewise linear function,  $h(E) = \lbrace E/\alpha, E \leq E_\mathrm{max}, E \geq E_\mathrm{max} \rbrace $, with $E_\mathrm{max}=30000$, $\alpha=30000$, and $\lambda_\mathrm{durotaxis}=20$. {\it (D)} Gaussian function, $h(E) = \exp \left (-(E-E_0)^2/(2\gamma^2) \right )$, with $E_0=15000$ and $\gamma=2000$, $\lambda_\mathrm{durotaxis}=10$. Insets show typical cell shape for regions indicated by red bars. 
\bigskip

\noindent{\bf Video S1.} Behavior in silico of a single cell on substrates of 4 kPa, 12 kPa, and 32 kPa, for a duration
of 500 MCS per simulation. Parameter settings as in Figure 3.
\bigskip

\noindent{\bf Video S2.} Pairwise cell-cell interactions in silico on substrates of 4 kPa, 12 kPa, and 32 kPa, for a
duration of 500 MCS per simulation. Parameter settings as in Figure 4.
\bigskip

\noindent{\bf Video S3.} Network formation in silico on a substrate of 10kPa, for a duration of 3000 MCS. Video
represents a $0.75\times0.75\;\mathrm{mm^2}$ area ($300\times300$ pixels) initiated with 450 cells. Parameter settings
are as in Figure 5.
\bigskip

\noindent{\bf Video S4.} Network formation of bovine aortic endothelial cells on a 2.5 kPa polyacrylamide gel functionalized with RGD-peptide. Time lapse images were captured in 30 minute intervals over an 8 hour time period. Image size as in Figure~\ref{fig:networks}~{\it D}.
\bigskip

\noindent{\bf Video S5.} Sprouting in silico from a spheroid on a substrate of 10kPa, for a duration of 3000 MCS.
Video represents a $0.75\times0.75\;\mathrm{mm^2}$ area ($300\times300$ pixels) initiated with 450 cells. Parameter
settings are as in Figure 6. 
\bigskip

\noindent{\bf Supporting Text S1.} Documentation of C and Matlab code used for the simulations, including a detailed description of the finite-element model.
\bigskip

\noindent{\bf Protocol S1.} C and Matlab source code used for the simulations.
\bigskip


\end{document}